\documentclass[final,5p,times,twocolumn]{elsarticle}
\usepackage{amssymb, amsmath, amsbsy}
\usepackage{algorithmic, algorithm}
\usepackage[T1]{fontenc}
\usepackage{color}
\journal{Computers \& Fluids}

\newcommand{\clear}[1]{ }

\begin{document}
\begin{frontmatter}

\title{Wall Orientation and Shear Stress in the Lattice Boltzmann Model} 

\author[ift]{Maciej Matyka}
\author[ift]{Zbigniew Koza}
\author[iipwr,vra]{{\L}ukasz Miros{\l}aw}
\address[ift]{Faculty of Physics and Astronomy, University of Wroc{\l}aw, pl.\ M.\ Borna 9, 50-205 Wroc{\l}aw, Poland}
\address[iipwr]{Institute of Informatics, Wroc{\l}aw University of Technology, Poland}
\address[vra]{Vratis Ltd., Wroc{\l}aw, Muchoborska 18, Poland}

\begin{abstract}
The wall shear stress is a quantity of profound importance
for clinical diagnosis of artery diseases. The lattice Boltzmann 
is an easily parallelizable numerical method 
of solving the flow problems, but it suffers 
from errors of the velocity field near the 
boundaries which leads to errors in the wall shear stress
and normal vectors computed from the velocity.
In this work we present a simple formula to calculate the wall shear 
stress in the lattice Boltzmann model and propose to compute
wall normals, which are necessary to compute the wall
shear stress, by taking the weighted mean over boundary facets lying
in a vicinity of a wall element. We carry out several tests and 
observe an increase of accuracy 
of computed normal vectors over other methods in two and three dimensions. 
Using the scheme we compute the wall shear stress
in an inclined and bent channel fluid flow and show 
a minor
influence of the normal on the numerical error, implying that  
that the main error arises due to a corrupted velocity field near the staircase
boundary. 
Finally, we calculate the wall shear stress in the human
abdominal aorta in steady conditions using our method and compare the
results with a standard finite volume solver and
experimental data available in the literature.
Applications of
our ideas in a simplified protocol for data preprocessing in medical applications are discussed.
\end{abstract}

\begin{keyword}
fluid flow \sep numerical methods \sep the lattice
Boltzmann method \sep LBM \sep wall shear stress \sep WSS 
\end{keyword}

\end{frontmatter}

\section{Introduction}
\label{Intro}

\begin{figure}
\centering
\includegraphics[width=0.44\textwidth]{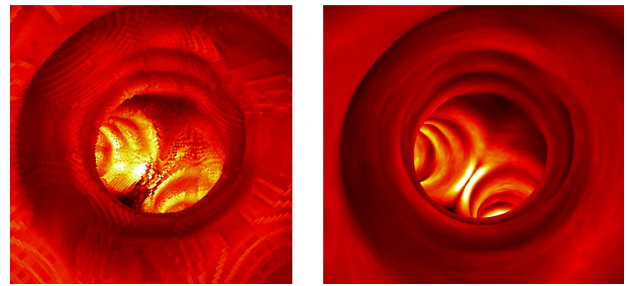}
\caption{(color online) Wall shear stress magnitude in the bifurcation
of the human abdominal aorta (the lighter color the
higher the stress) calculated using a lattice Boltzmann model (left)
and the finite volume method (right). \label{fig:acompare}}
\end{figure}

Computer simulations of the blood flow in 
the human circular system allow physicians 
to visualize the internal flow structure, a capability that 
might have a significant impact 
on diagnosis and medical treatment of arterial diseases (for review see \cite{Ku97,
Berger00, Anor10}). Digital simulations have been applied in many scenarios
and for various purposes e.g. to study the role of hemodynamic
forces in the development of 
atheromatous plaques in human cartoids atherosclerosis \cite{Dai04},
in a clinical study on the rupture risk assesment in the cerebral aneurysm \cite{Cebral04} 
or in a discovery of a shear-driven mechanism of platelet aggregation in arterial thrombosis \cite{Nesbitt09}. 
Direct results of hydrodynamical simulations (velocity and
pressure fields) are not practical in medical analysis.
To better understand the impact of the blood flow dynamics 
on the complex biochemical phenomena associated with the development of 
vascular diseases, a wall shear stress (WSS, see Fig.\ \ref{fig:acompare}) with its gradient (WSSG),
a hoop stress or an oscillatory shear index (OSI) are often computed. 
WSS is a tangential force 
exerted on the unit of the wall surface by the
flow \cite{Papaioannou05} and OSI describes its deviation  
from the average direction \cite{Ku85}.
WSS has been a subject of intensive research 
for many years, including studies on its role in development of human
atherosclerosis \cite{Shaaban00, Cecchi11},
the growth of an intracranial aneurysm \cite{Boussel08} and development
of the dissecting aneurysm in the human aorta \cite{Tse11}. 
WSS determines the structure, function and gene expression 
of an endothelial cells \cite{Dai04, Reneman06}.
It is now well established that a low shear stress 
is correlated with rapid development of various vascular malfunctions 
\cite{Caro09, Jou08, Carallo06, Meyerson01, Malek99}.
These facts motivate physicians to find applications of the WSS
in diagnosis \cite{Paszkowiak03,Pantos07} and to seek 
efficient and practical protocols for its measurement.

In order to compute WSS for patient-specific data
one has to acquire the organ geometry, convert and import it to
the Navier--Stokes solver, compute the velocity field and 
take gradients of its components at the wall. The process may be 
iterative and further time-averaging  might be necessary if the flow is
transient. The major difficulty of this procedure is the prohibitively long 
pre-processing and computation time which limits
its applicability in a real patient diagnosis. 
The pre-processing involves mainly the conversion of digital data 
from the computer tomography scans 
into a three-dimensional mesh   
required by standard finite element/finite volume solvers 
\cite{Milner98,Quarteroni00,Gohil11}. High
quality of the output mesh is required by flow solvers 
to work properly and computer-human interaction
during the meshing process is often necessary.  
The flow solver runtime depends on many factors 
e.g.\ solver type, mesh resolution,
flow properties and hardware used. It may be a bottleneck of the whole
procedure as well, particularly for geometries of high resolution, 
time dependent flows and problems involving fluid-structure interaction.
It is widely believed that the run time of 
fluid solvers could be significantly reduced by 
using efficient parallel algorithms on 
massively parallel architectures, e.g. graphics processing units 
(GPUs) \cite{Nickolls10}. 

Time issues mentioned above might be addressed with the lattice Boltzmann method
(LBM), a recently developed method for solving fluid flow problems
which is based on the kinetic theory of gases \cite{Aidun10}. The
method operates directly on
voxel data and thus does not require meshing at the level of accuracy 
required by standard solvers. It also proved excellent parallel scalability 
on supercomputers \cite{Pohl04,Peters10} as well as on emerging 
parallel architectures \cite{Harvey08,Peng08,Toelke08,Xian11}.
The method was validated against analytical solutions in a 
two-dimensional oscillatory channel flow and its second order
accuracy in space and time was confirmed \cite{He97,Cosgrove03}.
In medical applications, LBM was tested in solving time-dependend three-dimensional 
flows in models of the human
abdominal aorta \cite{Artoli06} and compared against finite element 
solvers in the superior mesentric artery blood flow \cite{Axner09}.
Non-Newtonian LBM models also exist e.g.\ the Carreau--Yasuda 
fluid model applied in cerebral aneurysm problem \cite{Bernsdorf09}.
Palabos, a parallel open source LBM implementation, 
was used to model blood flow in cerebral aneurysm and
compared to a finite volume flow solver
\cite{website:palabos,ChopardConf10} and the
difference between the velocity fields obtained in both methods 
was estimated to be around 10$\%$.
Application of LBM to modeling aneurysm and tumor development were recently discussed \cite{Chopard10}. 
He {\it et al.}\ compared LBM combined with the level-set method to an OpenFOAM solver in 
patient-specific geometry \cite{He09}. 

Despite the second order accuracy of the LBM stress 
tensor in bulk fluid being recently reported \cite{Kruger09,Kruger10}, 
the WSS in LBM suffers from inaccuracy introduced by a staircase boundary. The problem
may be eased with subgrid boundary conditions \cite{Boyd05}
or the recently developed unstructured mesh LBM formulation
\cite{Pontrelli11}, but both of them introduce additional
complexity into the solver code, making it more difficult to maintain
and parallelize. Recently Stahl {\it et al.}\ showed 
that a few lattice nodes away from the wall the error is significantly decreased in 
the standard LBM formulation \cite{Stahl10}. He suggests to calculate normal vectors
required by this procedure from the velocity
field as the fluid flowing along the noslip walls follows 
its geometry. Thus, the normal at the boundary is approximated by the
normal to the local velocity vector in a cell close to the
wall. However, this method behaves peculiarly bad in a vicinity
of the wall, with maximum error close to
90\%. Moreover, it is not suitable for time dependent flow
problems unless one solves an additional steady flow case. 
It also requires
a solution to the eigenvalue problem and does not provide the vector sense
relative to the wall. 

In this paper we discuss another way of computing the normal vectors
which does not suffer from above issues. If we look at the LBM 
staircase wall, we intuitively see its orientation. 
This is because we subconsciously average 
the input of all faces we see. 
A similar example is our brain's handling of a two-dimensional 
line on a computer display: if pixels are small enough, the line 
orientation is easily recognized by eye. We propose
to utilize similar averaging of a few boundary facets around the point
of interest to compute the normal vectors required for evaluating WSS.
The procedure is simple and its implementation consists of only a few
algebraic operations. 
Therefore, no eigenvalue problem has to be solved.
Additionally, the information about the normal sense is provided, and we show
that our procedure tends to provide more accurate results. 
Henceforth the normals calculated in this way will be called
`geometric normals' and the normals approximated from the velocity
field will be called `dynamic normals'.

The structure of the paper is as follows. In Sec.\ \ref{sec2} and
\ref{sec3} we briefly introduce the lattice Boltzmann method 
and the formula for the wall shear stress. In Sec.\ \ref{sec31} we explain 
how to compute geometric normals and verify them 
in simple geometries. The wall shear 
stresses in an inclined and a bent channel flow are calculated in 
Sec.\ \ref{sec4}. A comparison of WSS obtained with the 
lattice Boltzmann and a finite volume methods in the human abdominal
aorta is given in Sec.\ \ref{sec5}. Section \ref{sec6} presents the 
discussion of the results and Sec. \ref{sec7} is devoted to
conclusions.

\section{The lattice Boltzmann method}\label{sec2}

LBM is a computational method of solving 
the Navier--Stokes equations. It was developed mainly as a 
remedy to noise issues of the lattice gas 
automata (LGA) \cite{McNamara88, Higuera89}. In LGA the local state 
of the system is described by a boolean variable $c_\mathrm{i}$, where
$i$ runs over lattice vectors (vectors that link neighbour sites in the
lattice). 
In contrast, LBM replaces $c_\mathrm{i}$ by a particle distribution function
$f_\mathrm{i}\in [0,1]$, which is interpreted as the probability that a particle found in 
the lattice node moves along the $i$-th direction.
A typical LBM algorithm runs two steps, propagation and 
collision, and is expressed by a set of discrete equations:
\begin{equation} \label{eq:lbmalgorithm}
f_\mathrm{\alpha}(\mathbf{x}+\mathbf{c}_\mathrm{\alpha}\delta t,t+\delta t) =
f_\mathrm{\alpha}(\mathbf{x},t)+\Omega_\alpha(\mathbf{f}),
\end{equation}
where $\mathbf{c}_\alpha$ is the $\alpha$-th lattice vector, $\delta
t$ is the time
interval (usually $\delta t=1$) and $\Omega_\alpha(\mathbf{f})$ is a
collision operator.
The collision operator may realize either a linear relaxation to the equilibrium
distribution function or the multirelaxation
of separate hydrodynamical and kinetic moments \cite{dHumieres92}. The
Chapman-Enskog analysis cane be used to show
that the model reproduces incompressible Navier--Stokes equations \cite{Luo00}.

As an implementation of LBM we used the Sailfish library \cite{website:sailfish}.
Sailfish works on GPU processors. It is written in Python and 
utilizes template-based methodology for
automatic device code generation on CUDA and OpenCL GPU platforms. 
The original code is freely available under the LGPL licence.

\section{The wall shear stress}\label{sec3}

When fluid flows over a rigid surface, the velocity
at the wall vanishes and the no slip boundary condition holds.
However, in a vicinity of the wall the tangential component 
of the velocity does not vanish; the corresponding gradient of the tangential velocity
component along the wall normal generates the wall shear stress, a force that is 
exerted by the fluid
on the wall's surface. WSS may be expressed as the difference between
the Cauchy stress
on a plane and its projection on the plane normal (see
\ref{app:detailswss}). For an incompressible, Newtonian 
fluid the WSS in the lattice Boltzmann method can be calculated in
terms of the non-equilibrium distribution function
$f_\alpha^{neq}$:
\begin{equation} \label{eq:finalwss}
\tau_i=\frac{\mu\omega}{c_s^2\varrho}f_\alpha^{neq} c_{\alpha j} n_j
\left( c_{\alpha i}-c_{\alpha k}n_in_k \right),
\end{equation}
where $c_s$ is the lattice sound speed, $\mu$ is the dynamic viscosity,
$\varrho$ is the fluid density, and $\omega$ is the LBM relaxation
parameter. $c_{\alpha x}$ and $n_x$ denote lattice and normal vector
components, i.e.
$n_i$ is the $i$-th component of the wall normal vector $\mathbf{n}$
and $c_{\alpha i}$ is the $i$-th component of the lattice vector
$\mathbf{c}_{\alpha}$. Einstein summation convention is implied.
A detailed derivation of Eq.
\ref{eq:finalwss} is given in \ref{app:detailswss} with its algorithmic
form in \ref{app:wssalgo}.

\subsection{Wall orientation}\label{sec31}

To calculate WSS with Eq.\ (\ref{eq:finalwss}) the wall orientation
and the corresponding wall normal $\mathbf{n}$ must be known.
A typical three-dimensional LBM simulation works on a discrete, 
uniform grid built of adjacent cubic cells.
If a fluid node has a no-slip boundary neighbour then
there is a boundary facet between them at which the
no-slip condition of zero velocity is fulfilled. 
The facet is perpendicular to the line
that joins node centers whereas its physical location depends on the
LBM relaxation time. 
For a three-dimensional model there are six 
possible boundary normals:
$\pm \mathbf{e_x} = (\pm 1,0,0)$, $\pm \mathbf{e_y} = (0,\pm 1,0)$, and
$\pm \mathbf{e_z} = (0, 0, \pm 1)$.

%
\begin{figure}
\centering
\includegraphics[scale=0.7]{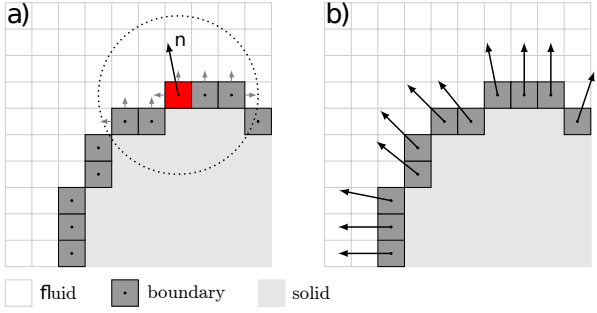}
\caption{(color online) (left) A geometric normal vector (long arrow denoted as $\mathbf{n}$) 
is calculated by averaging input from normals of neighbouring
fluid/no-slip boundary nodes (short arrows). Right: normal vectors 
computed for each individual
boundary node. Geometric normals are enlarged for 
easier viewing.\label{fig:normalfacet}}
\end{figure}

In order to calculate $\mathbf{n}$ we suggest taking an average
of normal vectors of the individual fluid/no-slip boundary facets
in a spherical vicinity of the cell. For
this we find all boundary facets whose centers lay in the sphere of
radius $r$ centered at a given cell (Fig.\ \ref{fig:normalfacet}a). 
If $\mathbf{b}_i$ is the boundary normal of the $i$-th facet then 
the above procedure may be expressed as the arithmetic mean:
\begin{equation}\label{eq:arithmetic}
\mathbf{n}=n_b^{-1}\sum_{i=0}^{n_b-1}\mathbf{b}_i,
\end{equation}
where $n_b$ is the number of facets within the neighborhood.
The same procedure is repeated for each boundary cell (Fig.\ \ref{fig:normalfacet}, right)

\subsection{Weighted scheme}

In the application of Eq.\ (\ref{eq:arithmetic}) an unacceptable 
averaging of normals that belong to separate or
opposite fragments of the object's surface may take place if $r$ is
taken to be too large.
In the hemodynamical scenario, this may occur in small arteries or
bifurcation regions where object radii are smaller that $r$ and
details may be lost due to averaging normals to opposite facets.
Moreover, if $r$ is too small, $\mathbf{n}$ is influenced by the 
staircase approximation of the surface and may change rapidly from
node to node. To deal with these issues we suggest to weight the mean
in Eq.\ (\ref{eq:arithmetic}):
\begin{equation}\label{eq:weighted}
\mathbf{n}=\left(\sum_i w_i\right)^{-1} \sum_{i}w_i\mathbf{b}_i,
\end{equation}
where the sum goes over neighour facets. We
choose $w_i=1/(1+d_i)^{\gamma}$, where  
$d_i$ is the distance between the center of the node for which
the average is computed and the center of the node with normal $\mathbf{b}_i$.

The exponent $\gamma$ defines how strongly concentrated the weighting
is, with $\gamma = 0$ corresponding to uniform weights 
(reducing to Eq.\ (\ref{eq:arithmetic})); the strength 
of weighting increases with the value of $\gamma$.

\subsection{Numerical tests for normal vectors}\label{sec:ball}

We test the above procedure in three-dimensional ball and tube
geometries for which exact normal vectors are known. First, a 
boolean array of size $L^3$ is generated such that
it holds $0$ if the voxel (three dimensional pixel) center lays inside 
the object and $1$ otherwise. Boundary cells are represented by 
voxels equal to $1$ having at least one adjacent zero neighbour. 
The object (ball or tube) radius is $R=L/2$ and we test three
resolutions: $L=25^3, 50^3$ and $100^3$. 
\begin{figure}
\centering
\includegraphics[width=0.3\textwidth]{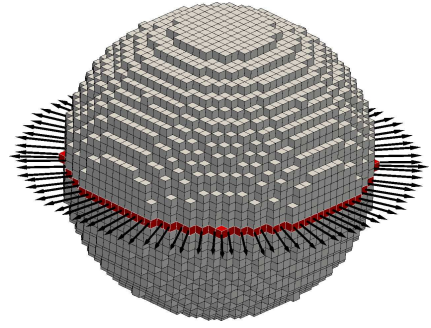}
\caption{(color online) Geometry of the ball for L=25 (lattice units). Normal vectors 
calculated at half vertical distance are computed with Eq.
(\ref{eq:weighted}) for $\gamma=1$. The averaging radius is
$r=4$ l.u.\label{fig:voxelball}}
\end{figure}
We calculate normals on the circle created by boundary nodes at half vertical
position. As an example, in Fig.\ \ref{fig:voxelball} the geometry of a ball for L=25 is 
visualized with normal vectors calculated using our procedure. 
\begin{figure}[!t]
\includegraphics[width=0.35\textwidth]{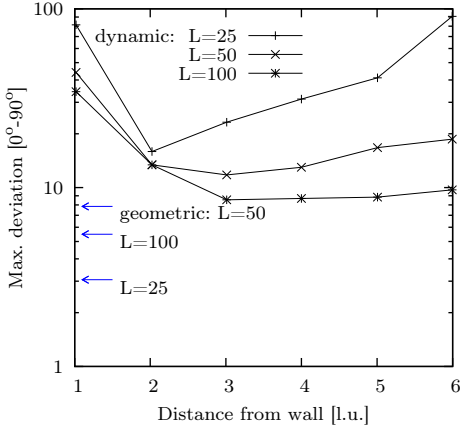}
\caption{
The maximum error of a normal vector (measured as the angle between
numerical and exact vectors) for two numerical methods in
a three-dimensional ball. 
Arrows represent geometric normals computed with Eq.\ (\ref{eq:weighted})
using $\gamma=1$ and $r=4$.\label{fig:ballplot}}
\end{figure}
Fig.\ \ref{fig:ballplot} depicts the maximum
error between geomeric and exact normals. Results are compared
to dynamic normals at various distances
from the wall \cite{Stahl10}.
Geometric normals are found to be more accurate.
\begin{figure}[!t]
\includegraphics[width=0.35\textwidth]{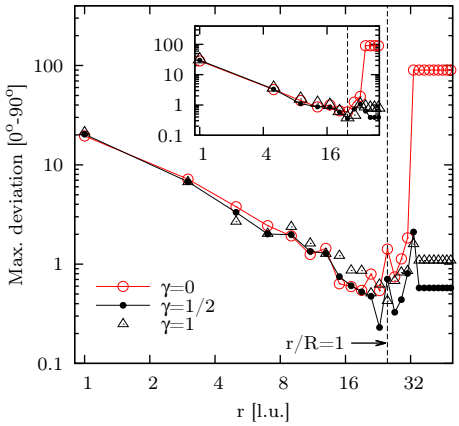}
\caption{(color online) The maximum error of geometric normal
(\ref{eq:weighted}) for three values of $\gamma$ as a function of the averaging radius $r$
in a three-dimensionall ball and tube (inset) with radii $R$=25. \label{fig:rave}}
\end{figure}
To test the influence of the averaging radius on the accuracy of 
the normal for various exponents $\gamma$ we vary $r$ and calculate the
error on a ball and a tube of radii $R$. The resulting error is
a non-monotonous function of $r$ with a minimum at 
at $r/R=1$ (see Fig. \ref{fig:rave}). At $r/R>1$ the mean error for $\gamma=0$  
grows rapidly whereas for $\gamma\neq 0$ it remains constant (with
slight favour to $\gamma=1/2$). A large error for $r/R>1$ for $\gamma=0$
comes from the fact that the averaging area covers the entire object, the 
resulting normal is $0$, and the information on the object 
geometry vanishes. If $\gamma\neq 0$, the problem is avoided because 
normal vectors defined on facets on the boundary lying on the opposite
side of the object have much lower weights. 

\section{Wall shear stress in a channel flow} \label{sec4}

In this section we test the accuracy of the
WSS equation (\ref{eq:finalwss}) combined with geometric normals in 
an inclined and bent channel flows.

\subsection{Inclined channel}

A two-dimensional channel of an initial resolution $120\times 20$
lattice units (l.u.) was inclined at angle $\alpha$
to the $x$ axis so that the radii of the inlet and outlet  were kept constant. The
analytical velocity profile (Poiseuille flow) was imposed on the inlet and outlet
boundaries. The lattice viscosity was $\nu=0.01$ and the maximum velocity 
in the channel center was $u_{max}=0.01$ (lattice units). Using
deviations from the 
analytical velocity field we found the steady state in less 
than $10^{4}$ iterations. We ran the 
flow for $\tan \alpha\in[0;2]$ and calculated dynamical and
geometrical normal vectors in a strip of length
$80$ l.u. (centered horizontally).
\begin{figure}
\centering
\includegraphics[width=0.45\textwidth]{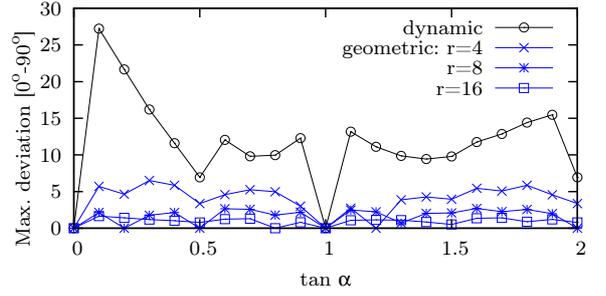}
\caption{
(color online) The maximum error of geometrical and 
dynamical normal in a channel at 
various inclination angles. For geometrical scheme three $r$ are tested.
\label{fig:normal1}}
\end{figure}
\begin{figure}
\centering
\includegraphics[width=0.45\textwidth]{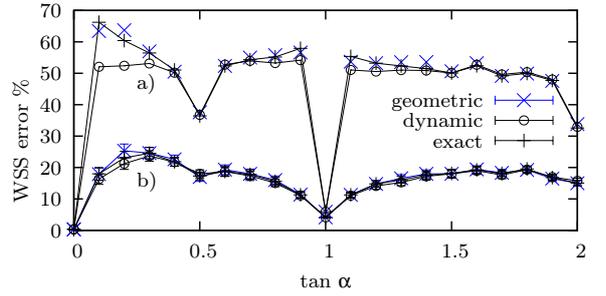}
\caption{
(color online) The maximum (a) and average (b) over $80$ neighbouring
sites of the error of WSS computed using various types of normal
vectors in a channel at various inclination angles. Error bars 
in the average represent the standard error.
\label{fig:inclinedwss}}
\end{figure}
In Fig.\ \ref{fig:normal1} the maximum angle between the numerical 
and exact vectors is given as a function of the channel 
inclination angle. The error of the geometric procedure
decreases with $r$ and is smaller than for the dynamic scheme. 
The most accurate results were obtained for $\tan \alpha=0$ and
$\tan \alpha=1$.

Next, we compute WSS using Eq.\ (\ref{eq:finalwss}) and show
that in an inclined channel WSS is essentially independent of 
the choice of the normal vector type (Fig.\ \ref{fig:inclinedwss}).
This implies that the largest contribution to the error comes from the
velocity field near the staircase boundary. This fact is also visible
in Fig.\ \ref{fig:normal1} where the large error of dynamic normals is
caused by the velocity field inaccuracy.
The maximum 
relative error of numerical WSS calculated at a single node 
is much larger than the error of the mean taken over 80 nodes. 
Therefore one could use an equation
similar to Eq.\ \ref{eq:weighted}, where a spatial
average of WSS in a vicinity of the wall element is computed: 
\begin{equation}\label{eq:auave}
\boldsymbol{\tau}=\left(\sum_i
w_i\right)^{-1}\sum_{i}w_i\boldsymbol{\tau}_i,
\end{equation}
where $w_i$ are weights of the same form as used for normals. However,
we leave applicability of Eq.\ (\ref{eq:auave}) to more complex flows 
as an open problem.

\subsection{Bent channel}

Next, we investigate the flow in a two-dimensional bent channel geometry 
defined by its inner and outer radii ($R_1$ and $R_2$, respectively). For the mesh of resolution $W\times H$ 
a site placed at $(i,j)$ is of fluid type if $R_1<|(i-W/2, j)|<R_2$,
otherwise it is a no-slip site (or it is excluded). We take $R_1=10$ and
$R_2=20$ and the mesh resolution $45\times 25$ l.u.\ and run the steady state 
simulation with $\nu=0.01$ and analytical inlet/outlet
boundary conditions.
\begin{figure}
\centering
\includegraphics[width=0.42\textwidth]{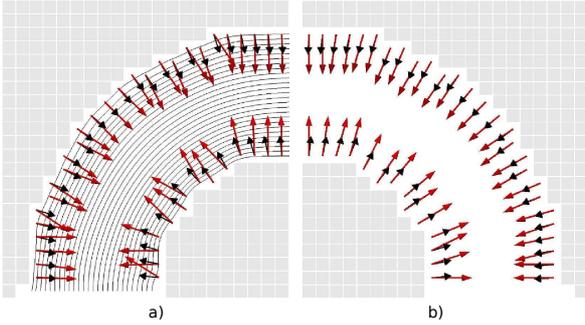}
\caption{(color online) Exact normals in a bent channel flow 
(short arrows) compared to dynamic (a) and geometric
(b) normals. Boundary cells are empty and only part 
of the channel is shown for both methods (the other parts are symmetric).\label{fig:bend1}}
\end{figure}

We calculated normals using both procedures and compare them 
visually against exact normals 
(Fig.\ \ref{fig:bend1}). To understand the influence of the flow field
near staircase wall on dynamic normals the flow field is shown 
using streamlines (Fig.\ \ref{fig:bend1} a). 
We determined the angle between each individual pair of 
numerical and exact vectors using the data from Fig.\ \ref{fig:bend1}
and found the maximum error of the dynamic scheme to be around 25$^{\circ}$. 
For the geometric scheme the error
is kept below 10$^{\circ}$. The average errors are calculated as 
$10.8^{\circ}$ and $3.9^{\circ}$ for dynamic and geometric vectors, respectively.
\begin{figure}
\centering
\includegraphics[width=0.47\textwidth]{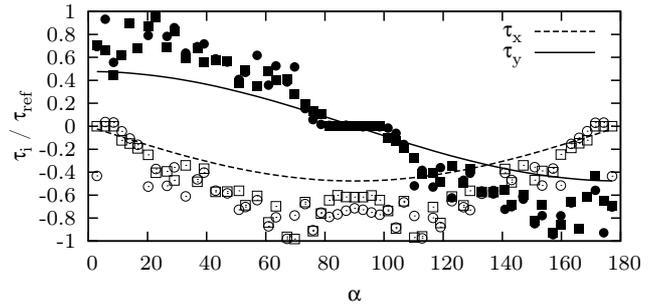}
\caption{WSS components in the bent channel flow computed with Eq.\
(\ref{eq:finalwss}) at various 
angular positions ($2$ l.u. from the inner channel wall). 
Results are obtained using geometric (squares) and dynamic (circles) normals.
Empty and filled symbols represent the $x$ and $y$ component, respectively.  
Data are normalized by $\tau_\mathrm{ref}=\max(\tau_i)$. 
\label{fig:bendwss}}
\end{figure}

Then we computed WSS at the inner wall at various angular positions
and plotted its components against exact solutions (Fig.\ 
\ref{fig:bendwss}). The numerical data
follow the analytical solutions. Regions of 
larger deviations may be recognized (e.g.\ $\tau_y$ at $\alpha \in (20,40)$) 
mostly because WSS was calulated at the first node next to the boundary. An interesting
observation may be done for the central region ($\alpha\in(80,100)$)
where introduction of the geometric normals slightly decreased the 
error for $\tau_x$ (similarly for $\tau_y$). The reason could be that the 
central part of the bent channel is horizontally flat in the grid 
representation (see Fig.\
\ref{fig:bend1}) and the velocity field follows the geometry. 
Therefore, the resulting dynamic normals are almost
perpendicular to the surface whereas they are not in the original
geometry (a bent inner circle). This results in a large systematic
deviation of the $x$ component of WSS which is slightly decreased if
more accurate geometric normals are used.

\section{A flow through the human abdominal aorta} \label{sec5}

Finally, we test the procedures described in previous sections in
a real situation and compute WSS in an abdominal aorta model based 
on patient-specific data. For this purpose we utilize data
from the Virtual Family (VF) \cite{Christ10} -- a freely available
library of high resolution anatomical human body models based on  
MRI measurements. We use Eq.\ (\ref{eq:finalwss}) to compute the geometric normals and Eq.\
(\ref{eq:arithmetic}) for WSS.

\subsection{Data preparation}\label{sec:datapreparation}\label{sec51}

From several models provided by the VF we chose Duke -- a 34 year 
old man. Using the software included in VF  we manually selected a
part of his abdominal aorta 17.8 cm over and 5.2 cm under the bifurcation
point. We exported voxel data as .raw files that encode a tissue type with integers. 
Then we used the CVMLCPP library \cite{website:cvmlcpp}
ray marching algorithm to approximate the artery surface. The
resulting mesh was of very poor quality, therefore further smoothing was
done using MeshLab \cite{website:meshlab}. At this point we either
voxelized the mesh to get various grid resolutions for the LBM or
continued meshing until we obtained a three-dimensional
unstructured volumetric mesh using NETGEN \cite{website:netgen}. Next
we applied the boundary conditions and exported the mesh in
the neutral NETGEN format for further use in FVM software.

\subsection{Simulation}

First, we ran the steady state LBM simulation using the Sailfish library 
adjusted for our purposes. We imposed a flat velocity 
profile on the inlet and zero pressure at both outlets of the model.
Due to the tortuous geometry of the aorta the 
flat inlet velocity causes a macroscopic flow with a slightly lower effective
flowrate. 
We used the following parameters: $u_{0,p}=0.0764$ m/s (the velocity at inlet), $l_{0,p}=0.03$ m (the
diameter of the inlet), $\nu=3\cdot 10^{-6}$ m$^2$/s (kinematic viscosity of the fluid),
$\varrho_p=1000$ kg/m$^3$ (the fluid density). Thus, the
characteristic time 
$t_{0,p}=l_{0,p}/u_{0,p}=0.393$ s and the Reynolds number Re=300
(based on an effective flowrate). The
model was subdivided into $461\times 108 \times 163$ voxels ($5$mm
resolution). The number
of grid nodes along the inlet diameter was $N$=60, thus $\delta x=0.0166$. We chosed
$u_{lb}=0.125$, for which $\delta t=u_{lb}\cdot\delta x=0.00208$ was
found. The final lattice viscosity
$\nu_{lb}=\delta t/\delta x^2 \cdot 1/\mathrm{Re}=0.0098$ was found. We
continued the simulation until the steady stade was reached 
(41~000 time steps) based on the observation of the velocity and
pressure field convergence. The resulting distribution
function that describes the steady state was used in Eq.\
(\ref{eq:finalwss})
to get WSS at $2$ l.u.\ away from the boundary. We used our geometric
normals in Eq.\ (\ref{eq:finalwss}) as well as to move away from the
boundary nodes. We first computed a dimensionless counterpart of $\tau_{lb}$, namely
$\tau_{d}=\varrho_{lb}\cdot \delta x^2/\delta t^2 \cdot
\tau_{lb}=63.7\tau_{lb}$. Then, the stress in [Pa] was obtained using
$\tau_p=\varrho_p\cdot l_{0,p}/t_{0,p}\cdot \tau_d=5.84\tau_d$.
\begin{figure*}[!ht]
\centering
\includegraphics[width=0.75\textwidth]{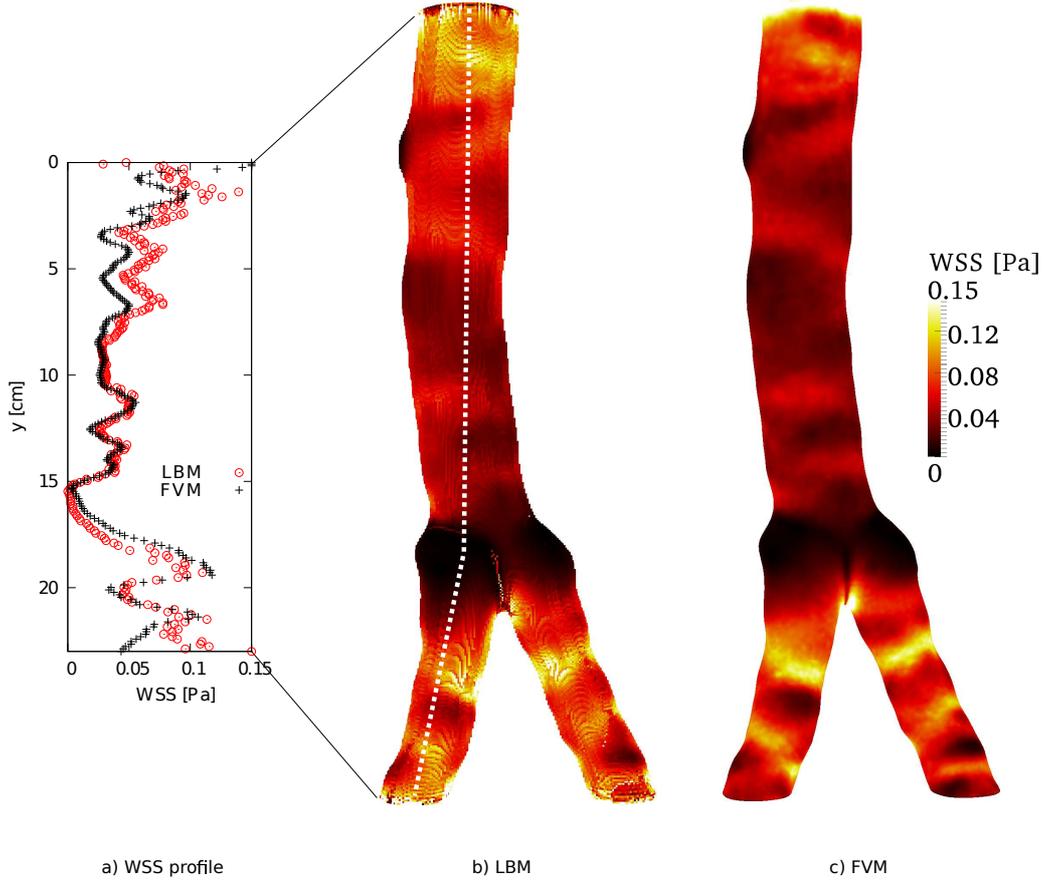}
\caption{(color online) Wall shear stress calculated using LBM and FVM. 
The lighter the color, the larger WSS acts on the
surface. 
\label{fig:acompare3}}
\end{figure*}

Second, we ran the same steady flow using finite volume Semi Implicit Method for
Pressure Linked Equations (SIMPLE) \cite{Patankar80}. We used simpleFoam
solver from OpenFOAM \cite{website:openfoam} with
previously generated mesh built of 496~656 tetrahedral elements. We 
applied a constant flowrate condition at the inlet with the flowrate 
Q=$2.1206\cdot 10^{-5}$ [m$^3$/s] corresponding to $Re=300$. The zero
pressure and zero velocity gradient was applied at both outlets.
We stopped the simulation after 1 second and confirmed that the convergence criteria was met (the initial and final residuals of
pressure and velocity was smaller than $10^{-5}$). Then, we computed WSS using
standard OpenFOAM tools. 

The resulting WSS in the abdominal aorta model computed with both
methods is visualized as a color map (Fig.\ \ref{fig:acompare} and Fig.\ \ref{fig:acompare3}).
The resulting ranges of WSS were $(6.74\cdot10^{-5}, 0.32)$ for LBM and
$(0,0.42)$ for FVM. We found that the discrepancy in maximum values is due to the
extreme values at the very boundary conditions, thus we encoded the
WSS magnitude in the range of WSS $\in(0,0.15)$ [Pa] only for better
visualization. The profile of WSS (Fig.\ \ref{fig:acompare3} a) is taken along the white 
dashed line (see Fig.\ \ref{fig:acompare3} b) from a gray scale 
projection of the results onto a plane.

The peak value in the bifurcation area is around $0.12$ [Pa] as 
observed in two branches behind
the bifurcation point where the wavy profile of WSS may be easily
recognized. The wavy profile of WSS seems to be mainly an effect 
of the geometry: the smaller profile diameter,  
the larger WSS. The quantitative comparison of WSS between 
LBM and WSS (Fig.\ \ref{fig:acompare3} a)
shows an excellent agreement in the area not influenced by boundary
conditions.

Next, we compare the WSS magnitudes obtained for both methods with 
the experimental data measured using the laser
photochromic dye tracer technique \cite{Bonert03}. There, for 
similar conditions (the abdominal aorta of the healthy 35-year-old
subject at rest and the steady flow) and the Reynolds number $Re=227$
(based on the inlet flowrate) the measurement values of WSS were in
the range $\tau\in(0.06,0.3)$ [Pa]. We found our numerical computations 
agree well with the experimental data. The remaining difference in 
the minimum value of WSS is most probably an effect of the
patient-specific geometry as we found the minimum of WSS in
Fig.\ \ref{fig:acompare3}a at $y=16$cm, where the enlargement of the
cross-section of our aorta model is clearly noticeable.

\section{Discussion}\label{sec6}

Comparison of normal vectors computed for three-dimensional objects
with two different methods given in
Sec.\ \ref{sec31} indicates an improvement in accuracy of our
procedure against the dynamic scheme based on the velocity field. The improvement is 
significant e.g.\ close to the wall for a small lattice 
the error for dynamic normals is close to $90^\circ$ but it remains
below $10^\circ$ in our method (Fig.\ref{fig:normal1}). 
Additional tests in the flow through an inclined (Fig.\ \ref{fig:normal1}) 
and bent (Fig.\ \ref{fig:bend1}) channel flows 
confirm higher accuracy of geometric normals. 
Application of the geometric normals gives practically the same WSS 
as when its dynamic or exact 
counterparts are used in an inclined channel. In the bent channel,
however, some differences are visible if separate stress
components are analyzed (see Fig.\ \ref{fig:bendwss}). We found a region
of an increased WSS accuracy based on the geometric normals 
(central part of the inner wall).
We believe this observations favorize 
the use of our scheme as it is simpler, more efficient 
(fewer algebraic operations) and provides the sense of the normal.
We are aware that our method is nonlocal which
may lead to complications in parallel implementation, this, however, might be 
solved e.g.\ by using algorithms that utilize fast shared memory of GPU 
processors.

Our tests in a real patient-based geometry show an excellent agreement 
with a standard finite volume solver. The WSS in
the fluid flow through the abdominal aorta is both qualitatively and quantitatively 
the same in LBM and FVM (see Fig.\ \ref{fig:acompare3}).
The slight difference between both methods at inlets and
outlets is an effect of differences in the boundary conditions used.
The data in Fig.\ \ref{fig:acompare3} a) give an impression of what is a typical
length scale at which the imposed inlet boundary conditions play a role in
hemodynamical simulations (e.g. we got around $10$ cm from the inlet for the given
setup). 

To get acceptable matching of the results in Fig.\ \ref{fig:acompare3} we moved 
the computation 2 nodes away from the wall (as suggested in \cite{Stahl10}).
This procedure has, however, some limitations. First, while producing color-maps of WSS we noticed some missing
nodes (see discontinuity in Fig.\ \ref{fig:acompare3}b just above the
bifurcation point). The reason for it is the following: if we start from two neighbouring sites and
move some distance along their normals, the destination
nodes no longer have to be neighbours (especially if normals were pointing in
different directions) and empty nodes at the surface may be visible.
This artifact does not influence WSS values obtained. To produce
decent visualization one could additionaly compute WSS for
those empty sites, e.g. in the Fig.\ \ref{fig:acompare} we visualize WSS
at a distances d=2 and at d=0 which practically eliminated the
problem of empty sites.
Second, moving away from boundary may be unwanted in general because one could argue the quantity is no longer the wall
shear stress but rather a near-the-wall shear stress. 
To make the use of the information at the wall
surface only, we suggested averaging of WSS at the 
wall in the neighbourhood of a site for which WSS is calculated (see
Eq.\ \ref{eq:auave}). Averaging decreased the error of
WSS at almost each site of the inclined channel flow by more than 50\%
(see Fig.\ \ref{fig:inclinedwss}).
To make the WSS averaging procedure a part of a complete protocol one should further test 
its behavior for smaller averaging radii and various
three-dimensional geometries.

We found that for a single model the preparation of 
FVM mesh described in Sec.\ \ref{sec51} takes around an hour without  
data segmentation (the Virtual Family implements this). 
In contrast, LBM could utilize the segmented MRI/CT data directly and skip 
the meshing process. We can imagine that, if the gray level was decoded
correctly into some of physical variables used in LBM simulation (e.g.\
endothelial wall roughness) then these data could be used directly
without segmentation. In this way one could turn the largest weakness of
the LBM method (staircase approximation of the boundary) into its
strongest point in hemodynamical applications 
(no need for tedious data preprocessing), where our simple procedure
for normals would be of high importance.

\section{Conclusions}\label{sec7}

There are several reasons why one would want to calculate the wall
orientation in LBM using the geometric procedure described here.
First, our method gives a gain in accuracy over dynamic  
normals. Second, the computation overhead is much smaller as the
averaging is done using a few simple algebraic operations 
rather than by solving an eigenvalue problem.
Third, we showed that our procedure gives practically the same WSS as if 
exact normals were used. Fourth, the computation of geometric normals is
-- by definition -- independent of the Reynolds number whereas dynamic
normals are dependent on Re, which is unphysical. Fifth, it is
suitable for the time dependent flows. Sixth, the given
scheme provides the sense of normals that might be crucial in the 
calculations that require determination of separated wall shear stress 
components e.g. oscillating shear index and wall shear stress gradient.

\section{Acknowledgement}
The publication has been prepared as part of the  project of the City
of  Wroc{\l}aw, entitled -- ``Green Transfer'' -- academia-to-business
knowledge transfer project co-financed by the European Union under the
European Social Fund, under the Operational Programme Human Capital
(OP HC): sub-measure 8.2.1 (MM). ZK was supported by MNiSW grant No.\ N
519 437939. We acknowledge support and discussions
from Dominik Szczerba, Jakub Kominiarczuk, Ziemowit Malecha, Micha{\l}
Januszewski, Jonas Latt and Bastien Chopard. Additionaly to the software 
mentioned in the text we used also Palabos, 
Paraview and Techdig. We gratefully acknowledge hardware donation from
Nvidia.

\appendix
\section{The stress on arbitrary plane in LBM}\label{app:detailswss}
The stress tensor in fluids is a sum of pressure 
and viscous terms:
\begin{equation}\label{eq:stresstensor}
\sigma_{ij}=-p\delta_{ij}+\sigma_{ij}'.
\end{equation}
\begin{figure}
\centering
\includegraphics[scale=0.9]{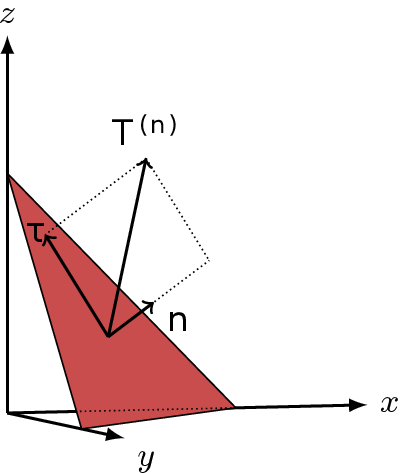}
\caption{(color online) The overall stress $\mathbf{T}^{(n)}$ and its tangential
part $\boldsymbol{\tau}$.\label{fig:wss}}
\end{figure}
The Cauchy formula gives the overall stress 
on the wall with a normal vector $\mathbf{n}$ (Fig. \ref{fig:wss}):
\begin{equation}\label{eq:cauchy}
T^{(\mathbf{n})}_{i}=\sigma_{ij}n_j.
\end{equation}
The shear stress may be computed as the difference between the overall stress
and its projection onto the normal:
\begin{equation}\label{eq:wsslast1}
\tau_i = \sigma_{ij}n_j-(\sigma_{kj}n_jn_k)n_i.
\end{equation}
We insert Eq. (\ref{eq:stresstensor}) into Eq. (\ref{eq:wsslast1}),
and since $\delta_{jk}n_jn_k=1$, the equation for WSS reads:
\begin{equation}\label{eq:wsslast2}
\tau_i = \sigma'_{ij}n_j-(\sigma'_{kj}n_jn_k)n_i.
\end{equation}
For the incompressible, Newtonian fluid
the viscous stress $\sigma'_{ij}$ is a linear function of 
the rate of strain $\varepsilon_{ij}$:
\begin{equation}\label{eq:stressshear}
\sigma_{ij}'= 2\mu \varepsilon_{ij}.
\end{equation}
%
Combination of Eq. (\ref{eq:wsslast2}) and Eq. (\ref{eq:stressshear})
gives:
\begin{equation}\label{eq:taustrain}
\tau_i = 2\mu
\left(\varepsilon_{ij}n_j-\left(\varepsilon_{kj}n_jn_k\right)n_i\right).
\end{equation}
A definition of the rate of strain tensor in LBM, which
uses the non-equilibrium part of the distribution function reads:
\begin{equation}\label{eq:strainlbm}
\varepsilon_{ij}=\frac{\omega}{2c_s^2\varrho}\sum_\alpha f_\alpha^{neq} c_{\alpha i}c_{\alpha j},
\end{equation}
where $f_\alpha^{neq}=f_\alpha-f_\alpha^{eq}$ and $f_\alpha^{eq}$ is
an equilibrium Maxwell-Boltzmann distribution. Finally, we combine 
Eqs.\ (\ref{eq:taustrain}) and (\ref{eq:strainlbm}) and and get Eq. (\ref{eq:finalwss}).
\section{WSS algorithm} \label{app:wssalgo}

\begin{algorithm}
\begin{algorithmic}
\STATE $dim \gets 3$ 
\STATE $C \gets (\mu\omega)/(c_s^2\varrho)$
\FOR{$i = 0 \to dim-1$} 
\STATE $sum\alpha \gets 0$
 \FORALL{$\alpha$} 
  \STATE $sumk \gets 0$
   \FOR{$k = 0 \to dim-1$}
   \STATE $sumk \gets sumk + c_{\alpha k}n_in_k$
  \ENDFOR
  \STATE $sumj \gets 0$
   \FOR{$j = 0 \to dim-1$}
     \STATE $sumj \gets sumj + c_{\alpha j} n_j \left( c_{\alpha i}- sumk \right)$
   \ENDFOR
   \STATE $f_\alpha^{neq} \gets f_\alpha-f_\alpha^{eq}$
   \STATE $sum\alpha \gets sum\alpha + f_\alpha^{neq} \cdot sumj$
 \ENDFOR
 \STATE $\tau_i \gets C \cdot sum\alpha$
\ENDFOR
\STATE $WSS \gets |\boldsymbol \tau|$
\end{algorithmic}
\caption{Calculation of WSS from Eq. (\ref{eq:finalwss}).}
\end{algorithm}

\bibliographystyle{elsarticle-num-names}
\bibliography{aorta}

\end{document}